\documentclass[sigconf,natbib=true]{acmart}
\usepackage{amsmath}

\usepackage{graphicx}
\usepackage{natbib}
\usepackage{multirow} 
\usepackage{algorithm}
\usepackage{algpseudocode}
\usepackage{amsmath}
\usepackage[utf8]{inputenc}
\usepackage{soul}
\AtBeginDocument{%
  \providecommand\BibTeX{{%
    \normalfont B\kern-0.5em{\scshape i\kern-0.25em b}\kern-0.8em\TeX}}}

\copyrightyear{2025}
\acmYear{2025}
\setcopyright{cc}
\setcctype{by}
\acmConference[SIGIR '25]{Proceedings of the 48th International ACM SIGIR Conference on Research and Development in Information Retrieval}{July 13--18, 2025}{Padua, Italy}
\acmBooktitle{Proceedings of the 48th International ACM SIGIR Conference on Research and Development in Information Retrieval (SIGIR '25), July 13--18, 2025, Padua, Italy}
\acmDOI{10.1145/3726302.3730097}
\acmISBN{979-8-4007-1592-1/2025/07}

\makeatletter
\gdef\@copyrightpermission{
    \begin{minipage}{0.3\columnwidth}
        \href{https://creativecommons.org/licenses/by/4.0/}{\includegraphics[width=0.90\textwidth]{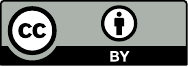}}
    \end{minipage}\hfill
    \begin{minipage}{0.7\columnwidth}
        \href{https://creativecommons.org/licenses/by/4.0/}{This work is licensed under a Creative Commons
        Attribution International 4.0 License.}
    \end{minipage}
\vspace{5pt}
}
\makeatother

\begin{document}


\title{Towards Brain Passage Retrieval}
\subtitle{An Investigation of EEG Query Representations}



\author{Niall Mcguire}
\email{niall.mcguire@strath.ac.uk}
\orcid{0009-0005-9738-047X}
\affiliation{
  \institution{NeuraSearch Laboratory\\University of Strathclyde}
  \city{Glasgow}
  \country{UK}}

\author{Yashar Moshfeghi}
\email{yashar.moshfeghi@strath.ac.uk}
\orcid{0000-0003-4186-1088}
\affiliation{%
  \institution{NeuraSearch Laboratory\\University of Strathclyde}
  \city{Glasgow}
  \country{UK}}
\renewcommand{\shortauthors}{McGuire and Moshfeghi}

\begin{abstract}
Information Retrieval (IR) systems primarily rely on users' ability to translate their internal information needs into (text) queries. However, this translation process is often uncertain and cognitively demanding, leading to queries that incompletely or inaccurately represent users' true needs. This challenge is particularly acute for users with ill-defined information needs or physical impairments that limit traditional text input, where the gap between cognitive intent and query expression becomes even more pronounced. Recent neuroscientific studies have explored Brain-Machine Interfaces (BMIs) as a potential solution, aiming to bridge the gap between users' cognitive semantics and their search intentions. However, current approaches attempting to decode explicit text queries from brain signals have shown limited effectiveness in learning robust brain-to-text representations, often failing to capture the nuanced semantic information present in brain patterns. To address these limitations, we propose BPR (\textit{\textbf{B}rain \textbf{P}assage \textbf{R}etrieval}), a novel framework that eliminates the need for intermediate query translation by enabling direct retrieval of relevant passages from users' brain signals. Our approach leverages dense retrieval architectures to map EEG signals and text passages into a shared semantic space. Through comprehensive experiments on the ZuCo dataset, we demonstrate that BPR achieves up to 8.81\% improvement in precision@5 over existing EEG-to-text baselines, while maintaining effectiveness across 30 participants. Our ablation studies reveal the critical role of hard negative sampling and specialised brain encoders in achieving robust cross-modal alignment. These results establish the viability of direct brain-to-passage retrieval and provide a foundation for developing more natural interfaces between users' cognitive states and IR systems.
\end{abstract}

\ccsdesc[300]{Information systems~Search interfaces}
\ccsdesc[500]{Information systems}
\ccsdesc[500]{Information systems~Information retrieval}
\ccsdesc[300]{Information systems~Users and interactive retrieval}

\keywords{Information Retrieval, EEG, Passage Retrieval, Query Representation, Brain Machine Interfaces}



\maketitle

\section{Introduction}
\label{introduction}
One of the fundamental challenges in Information Retrieval (IR) lies in bridging the gap between users' core information needs (INs) and their external formulation as (textual) queries~\cite{belkin1980anomalous, ingwersen1996cognitive, moshfeghi2016understanding}. The cognitive complexity of query formulation often results in queries that incompletely or inaccurately represent users' true needs, while physical constraints of input mechanisms can create additional barriers to effective information access~\cite{belkin1980anomalous, wolpaw2002brain}. This challenge is particularly acute for users with ill-defined information needs or physical impairments that limit traditional text input, where limitations in conventional interaction mechanisms can potentially exclude significant user populations from effective information access~\cite{wolpaw2002brain}. Foundational models of information-seeking behaviour illuminate the depth of this challenge. Taylor~\cite{taylor1968question} demonstrates that users must progress through distinct stages, from a visceral, unconscious information need to a compromised form suitable for an IR system, a progression that often results in significant information loss. Kuhlthau~\cite{kuhlthau2005information} extends this understanding by revealing how users struggle with uncertainty and vague thoughts during early search stages, precisely when they must articulate their needs most explicitly. The combination of these cognitive and physical barriers creates a fundamental gap between users' internal states and their ability to externalise these states through conventional interaction mechanisms~\cite{ingwersen1996cognitive, belkin1980anomalous}.

\begin{figure}
    \centering
    \includegraphics[width=0.9\linewidth]{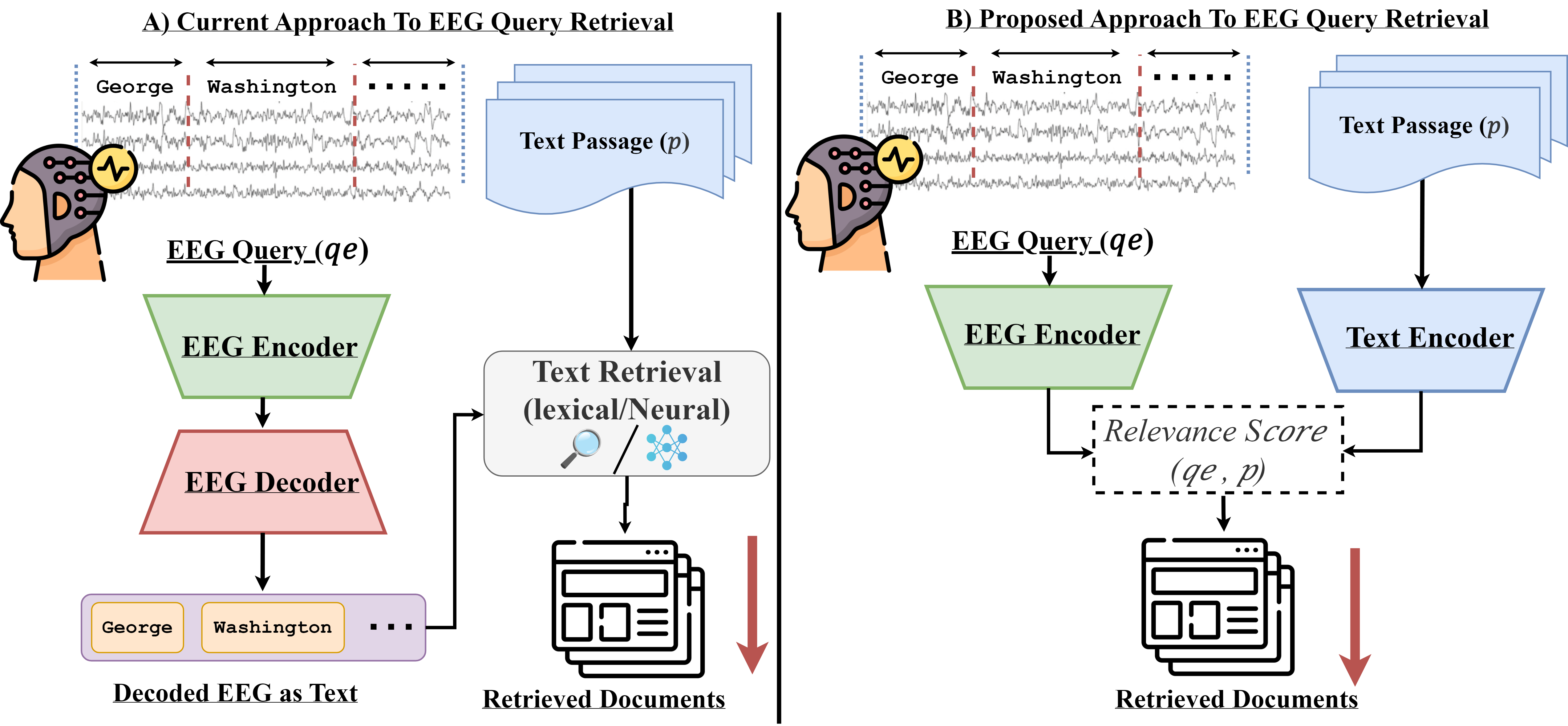}
    \caption{A) Traditional EEG-to-text pipeline requiring intermediate query decoding before retrieval. The approach first translates EEG signals into text queries before applying traditional (lexical/neural) text retrieval methods. B) Our proposed direct EEG query retrieval framework that eliminates the translation step by learning a shared embedding space between EEG signals and text passages, enabling direct relevance scoring between brain activity and documents.}
    \label{fig:EE2TextvsBPR}
\end{figure}

Recent neuroscientific advances have revealed promising approaches to bridging this gap through direct measurement of users' cognitive states during information interaction~\cite{moshfeghi2016understanding, eugster2016natural, gwizdka2019introduction}. Studies have demonstrated that core IR processes manifest as distinct brain signatures, from information need formation~\cite{moshfeghi2019towards, michalkova2022information} to relevance judgements~\cite{kauppi2015towards, allegretti2015relevance, pinkosova2020cortical} and search satisfaction~\cite{paisalnan2022neural}. These neurophysiological insights have motivated researchers to explore brain-based approaches for enhancing query formulation. Initial implementations focused on Steady-State Visually Evoked Potentials (SSVEP)\cite{wolpaw2002brain}, these approaches primarily changed the input mechanism rather than addressing the fundamental gap between need and expression. More recent work using functional Magnetic Resonance Imaging (fMRI) and Magnetoencephalography (MEG) has demonstrated increasingly sophisticated semantic decoding capabilities \cite{jayalath2024brain}, from mapping word-level representations~\cite{mitchell2008predicting} to reconstructing continuous narratives~\cite{huth2016natural} and generating natural language from brain patterns~\cite{ye2024generative}. While these advances validate the possibility of capturing information needs directly from brain signals, both fMRI and MEG have severe practical limitations for real-world IR applications, requiring immobile participants in specialised facilities~\cite{ye2024query, kauppi2015towards} and involving prohibitive equipment costs~\cite{allegretti2015relevance}.

Electroencephalography (EEG) offers a more practical path forward, providing high temporal resolution and mobility at a significantly lower cost~\cite{binnie1994electroencephalography, mcguire2023song, mcguire2023ensemble, kingphai2023channel}. Recent studies have demonstrated EEG's ability to detect relevance judgments~\cite{pinkosova2020cortical, allegretti2015relevance, gwizdka2017temporal, michalkova2024understanding, pinkosova2022revisiting}, satisfaction~\cite{paisalnan2022neural}, and enable direct document recommendation from brain signals during reading~\cite{eugster2016natural, zhang2024improving}. However, current EEG-based approaches for document retrieval require the translation of brain signals into textual queries~\cite{wang2022open, lamprou2024role} (see Figure \ref{fig:EE2TextvsBPR}), which recent studies have identified as ineffective for learning generalisable semantic representations from EEG and rely on the memorisation of training data and teacher forcing to generate a textual query~\cite{jo2024eeg}. As a solution to this problem, we present BPR (\textit{\textbf{B}rain \textbf{P}assage \textbf{R}etrieval}), a framework that eliminates this translation step entirely by directly mapping brain signals to dense passage representations. Rather than attempting to convert brain activity into text queries, BPR projects brain patterns into the same semantic space as passages, treating brain activity itself as a query representation (see Figure \ref{fig:EE2TextvsBPR}). Our approach builds on dense retrieval architectures~\cite{karpukhin2020dense} but adapts them for brain data through specialised EEG encoders and cross-modal negative sampling strategies.
Our contributions include:
\begin{itemize}
\item The development of BPR, a novel framework for direct EEG-to-passage retrieval that achieves a 8.81\% improvement in Precision@5 over existing approaches by eliminating intermediate text translation
\item An effective adaptation of dense retrieval architectures for brain signals, incorporating specialised EEG encoders and cross-modal negative sampling to learn effective representations
\item Empirical validation demonstrating the first successful semantic alignment between EEG signals and document representations, with comprehensive ablation studies across multiple evaluation settings
\end{itemize}
Our empirical results establish that direct brain-to-passage retrieval outperforms existing approaches and demonstrates the feasibility of meaningful semantic alignment between brain signals and text without intermediate translation steps. These findings represent a promising step toward IR systems that can detect and respond to information needs in their most visceral form. 

\section{Related Works}


\subsection{Neuroscience \& Information Retrieval}
Neuroscience has transformed our understanding of how users interact with information systems by revealing the brain mechanisms underlying core IR processes. Early fMRI studies provided crucial insights into the formation of information needs, with \citet{gwizdka2019introduction,moshfeghi2016understanding} identifying specific activation patterns in the posterior cingulate cortex that signal when users recognise gaps in their knowledge. This work was extended by \citet{michalkova2022information}, who characterised the metacognitive states that precede active search behaviour, demonstrating that information needs have distinct brain signatures before users can consciously articulate them. The neuroscientific basis of IR extends beyond information need formation. Using fMRI, \citet{moshfeghi2013understanding} mapped how relevance judgments manifest in brain activity, while following work with EEG by \citet{allegretti2015relevance} provided converging evidence that these judgments have consistent neurological correlates across users and contexts. \citet{gwizdka2017temporal} revealed how attention and cognitive load fluctuate throughout the search process, while \citet{ji2024characterizing} demonstrated how combining EEG with eye-tracking can reveal distinct patterns in information-seeking behaviour. Collectively, these studies established that fundamental IR concepts have reliable neurological signatures that emerge before explicit user actions.

These neurophysiological insights have driven the development of systems attempting to bridge the gap between internal information needs and explicit queries. Early applications focused on enhancing existing IR interfaces through neurological feedback, with \citet{eugster2014predicting} developing methods to detect term relevance from brain signals during search tasks. Moreover, \citet{mcguire2024prediction} achieved 90\% accuracy in predicting information need formation from EEG signals during question-answering tasks, suggesting the possibility of detecting search intentions before query formulation.

In parallel to these works, researchers also began exploring direct brain-machine interfaces (BMIs). SSVEP-based systems \cite{vialatte2010steady} enabled users to construct queries through brain signals by selecting characters from a virtual keyboard, with each key flickering at a unique frequency detectable in the visual brain-machine cortex. While \citet{chen2022web} achieved impressive accuracy (91\%) with this approach, these systems still required users to explicitly construct queries, maintaining rather than eliminating the translation burden between information need and query formulation. This limitation highlighted a crucial challenge: while neuroscience had revealed how information needs manifest in the brain, and while systems could detect these needs with high accuracy, the fundamental problem of translating internal cognitive states into effective queries remained unsolved. This realisation motivated a shift toward direct semantic decoding from brain activity, attempting to capture information needs in their most visceral form through explicit translation into query terms.

\subsection{Semantic Decoding of Brain Signals}
Recent advances in semantic decoding from fMRI scans have shown promising results in reconstructing natural language from brain activity. \citet{ye2024generative} demonstrated the semantic reconstruction of continuous language by mapping brain representations to semantic space using large language models, while \citet{ye2024query} showed applications for query enhancement in retrieval models. Magnetoencephalography (MEG) has provided complementary insights, with \citet{mitchell2008predicting} and \citet{kauppi2015towards} achieving semantic decoding with high temporal resolution. However, both technologies have significant practical limitations for real-world implementation into present IR methodologies. fMRI requires participants to remain motionless within the scanner bore, where minimal head movements can corrupt data quality \cite{moshfeghi2016understanding}. MEG systems similarly constrain movement and require magnetically shielded rooms \cite{kauppi2015towards}. 

These physical requirements, combined with high equipment costs, limit deployment in real-world IR applications. These constraints motivated researchers to investigate more practical recording methods, namely EEG (as discussed in Section \ref{introduction}). \citet{wang2022open} introduced EEG-to-Text (EEG2Text), formulating EEG decoding as a machine translation problem with an EEG encoder and pre-trained decoder (BART) optimised to minimise the cross-entropy loss between the predicted text output and true text labels (see Figure \ref{fig:EE2TextvsBPR}). This framework influenced subsequent works such as DeWave \cite{duan2023dewave}, which incorporated discrete tokens to improve decoding performance. However, recent analyses have identified methodological limitations in these translation approaches. \citet{jo2024eeg} highlighted that these models rely on teacher forcing \cite{hao2022teacher} during inference (a technique where the model uses ground truth tokens rather than its own predictions during sequence generation), leading to inflated performance metrics. Additionally, the models show similar performance with random noise input compared to actual EEG signals, suggesting that the model may learn to memorise dataset artefacts rather than meaningful brain-to-text mappings.

\begin{figure*}[h]
    \centering
    \includegraphics[width=\linewidth]{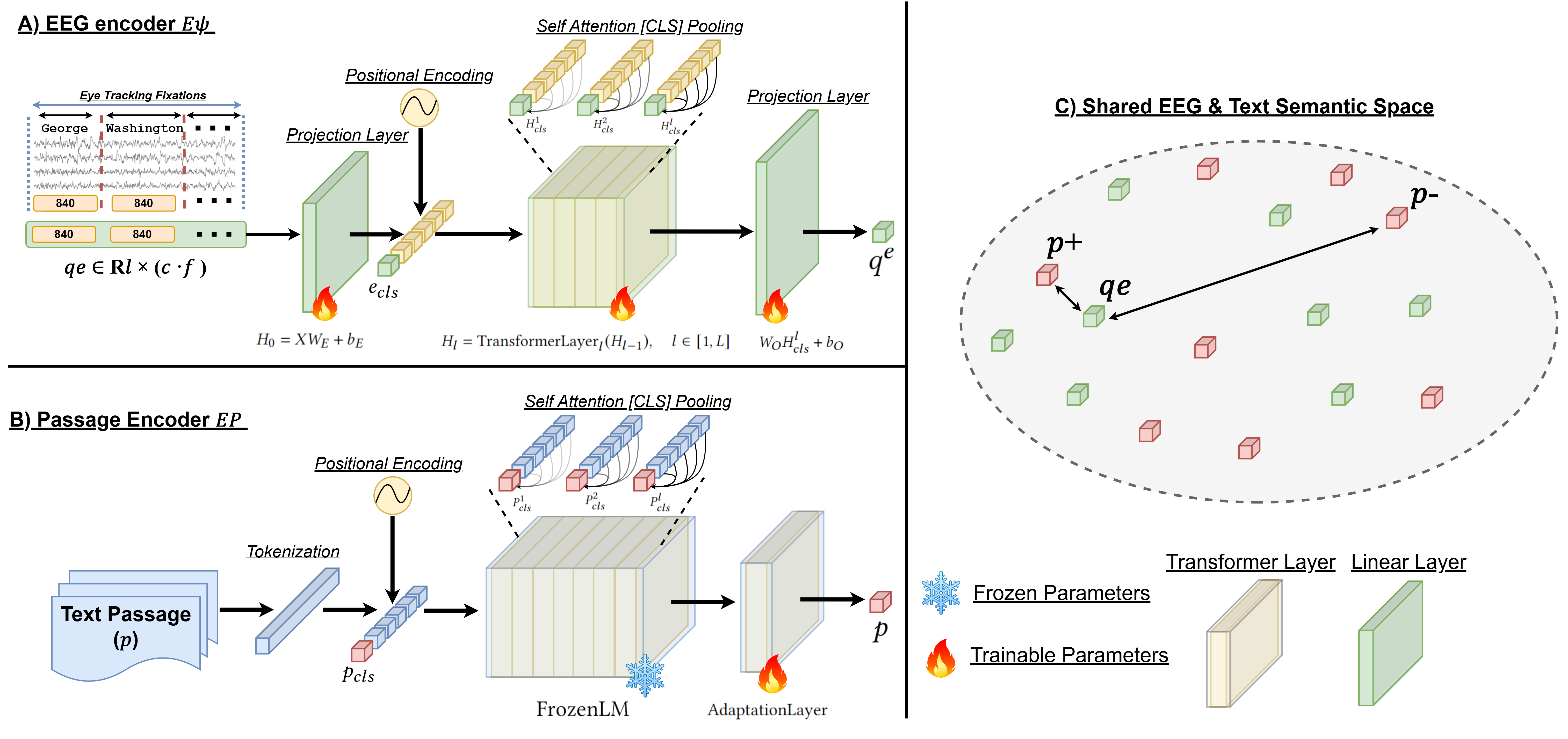}
    \caption{Overview of the BPR architecture. A) EEG encoder processes EEG signals recorded during naturalistic reading, using eye-tracking fixations to align EEG data with individual words. The architecture consists of initial projection layers, positional encoding, and transformer layers with self-attention mechanisms to generate EEG query representations. B) Passage encoder leverages a frozen pre-trained language model with an additional lightweight adaptation layer to generate passage representations. C) Shared semantic space visualisation demonstrating how EEG queries (green) and passages (red) are mapped into a common embedding space, where $p^+$ indicates positive passage matches and other red cubes represent negatives $p^-$. Flame icons indicate trainable parameters while snowflakes indicate frozen model components.}
    \label{fig:BPR-Overview}
\end{figure*}

Further work has explored the adaptation of speech models for EEG decoding. BrainEcho \cite{li2024brainecho} adapted the Whisper model for EEG-to-text conversion, but its reliance on parallel audio processing makes it unsuitable for comparison in naturalistic reading tasks. These methods also use smaller vocabularies than established EEG2Text studies \cite{duan2023dewave, wang2022open}, making direct comparisons impractical. The observed limitations of current EEG-to-text translation approaches highlight a fundamental challenge in facilitating passage retrieval using brain signals. Rather than pursuing further refinements to these translation methods, we propose eliminating the intermediate translation step entirely, and instead directly mapping EEG query representations to a shared semantic text space (see Figure \ref{fig:BPR-Overview}).

\section{Methodology}
\label{methodology}

\subsection{Task Formulation}
\label{task formulation}
While neural retrievers have significantly advanced the capabilities of IR systems through their ability to capture deep semantic relationships and contextual understanding beyond lexical matching \cite{karpukhin2020dense, xiong2020approximate, ren2021rocketqav2, khattab2020colbert}, they still rely on users translating their INs into explicit textual queries (as discussed in Section \ref{introduction}). This translation process presents fundamental challenges, as users often struggle to precisely specify their information needs \cite{belkin1980anomalous}, particularly when they are uncertain about what information would resolve their anomalous state of knowledge. Recent work in BMIs has demonstrated the feasibility of decoding semantic information directly from brain signals \cite{mitchell2008predicting, huth2016natural, caucheteux2022deep}. While these advances suggest promising directions for enhancing query formulation, current approaches face significant practical and technical limitations due to their demonstrated inability to learn robust brain-to-text representations \cite{jo2024eeg, wang2022open}. Rather than attempting to translate brain signals into explicit queries, we propose a direct brain-to-passage mapping approach that could better preserve the richness of users' cognitive states during reading while avoiding the complexities of intermediate translation steps.

Formally, let $D$ represent a passage as a sequence of $l$ words $\{w_1, w_2, ..., w_l\}$. As a subject reads this passage, their brain activity is recorded using an EEG device, producing a corresponding signal sequence $E = \{e_1, e_2, ..., e_l\}$. Each signal $e_i$ captures the neurological response when reading word $w_i$ through $c$ EEG channels and $f$ frequency bands, resulting in a feature vector $e_i \in \mathbb{R}^{c \cdot f}$. Given these EEG recordings and a corpus of $N$ passages $C = \{p_1, p_2, ..., p_N\}$, our goal is to retrieve the $k$ most relevant passages $R = \{p_{i_1}, p_{i_2}, ..., p_{i_k}\}$ in response to an EEG query $q^e \in \mathbb{R}^{l \times (c \cdot f)}$. This formulation extends traditional dense passage retrieval \cite{karpukhin2020dense} by replacing the text query encoder with an EEG query encoder. Our proposed model employs a dual-encoder architecture with the following components:
\begin{itemize}
\item An EEG encoder $E_{\psi} : q^e \rightarrow \mathbb{R}^d$ that projects brain signals to dense vectors
\item A passage encoder $E_p : p \rightarrow \mathbb{R}^d$ that maps text to the same embedding space
\end{itemize}

During inference, relevance between an EEG query and passage is computed using a similarity function $s$ that measures the semantic similarity between the encoded query and passage representations:
\begin{equation}
\text{score}(q^e, p) = s(E_{\psi}(q^e), E_p(p))
\end{equation}

The final retrieval process identifies the top-$k$ most relevant passages by maximising the similarity scores:
\begin{equation}
R = \underset{R' \subset C, |R'|=k}{\arg\max} \sum_{p \in R'} \text{score}(q^e, p)
\end{equation}

This framework enables direct mapping between brain signals and text passages (see Figure \ref{fig:BPR-Overview}) while maintaining the information present in brain activity during natural reading tasks.

\subsection{Dataset Creation}
\label{dataset creation}
Training neural retrieval models requires large quantities of query-document pairs \cite{karpukhin2020dense}, along with effective negative sampling strategies to create challenging contrasts during training. However, unlike traditional IR datasets \cite{nguyen2016ms}, EEG datasets of sufficient scale with predefined query-document pairs are not readily available. The ZuCo dataset \cite{hollenstein2018zuco, hollenstein2019zuco} was selected for this work as it represents one of the few publicly available large-scale EEG datasets with paired text recordings during natural reading, containing synchronised neurological signals and eye-tracking data from 1,000+ English sentences read by multiple participants. To address the limitation of predefined query-document pairs, we adapt the inverse cloze task (ICT) framework \cite{lee2019latent, chang2020pre} to construct synthetic training data from our EEG recordings. Following prior examples in dense retrieval \cite{karpukhin2020dense, xiong2020approximate, ren2021rocketqav2}, the ICT approach enables us to generate query-document pairs by treating spans of text as implicit queries while considering their surrounding context as relevant documents.

Formally, using the notation established earlier (where $D = \{w_1, w_2, \dots, w_l\}$ represents a passage and $E = \{e_1, e_2, \dots, e_l\}$ represents its corresponding EEG recordings), we extract a text span $S = \{w_i, w_{i+1}, \dots, w_{i+j}\}$ to serve as a pseudo-query. The EEG signals corresponding to this span $q^e = \{e_i, e_{i+1}, \dots, e_{i+j}\}$ form our query representation. With probability $p_{mask}$ (set to 0.9 in our implementation), we remove this span from the passage to form the positive document: $D^+ = \{w_1, \dots, w_{i-1}, w_{i+j+1}, \dots, w_l\}$. Otherwise, with probability $1 - p_{mask}$, we retain the query span in the document, creating a more challenging learning scenario where the model must learn robust matching strategies beyond exact token matching. Algorithm~\ref{alg:ict} details our ICT implementation. The algorithm takes as input the passage tokens, their corresponding EEG signals, query length ratio, and mask probability, then:

\begin{enumerate}
    \item Computes the query length as a fraction of the total document length.
    \item Randomly selects a starting position for the query span.
    \item Extracts both the text query and its corresponding EEG signals.
    \item Randomly decides whether to remove the query span from the document based on $p_{mask}$.
    \item Returns the EEG query and document pair for training.
\end{enumerate}

\begin{algorithm}[t]
    \caption{Inverse Cloze Test for Brain Query Generation}
    \label{alg:ict}
    \begin{algorithmic}[1]
        \Require{Document tokens $D = \{w_1, w_2, \dots, w_l\}$, EEG signals $E = \{e_1, e_2, \dots, e_l\}$, mask probability $p_{mask}$}
        \Ensure{EEG query $q^e$, modified document $D'$}
        \State $L \leftarrow \lfloor l \cdot 0.3 \rfloor$ \Comment{Set query length to 30\% of document}
        \State Select random index $i$ where $0 \leq i \leq l - L$
        \State $Q \leftarrow \{w_i, w_{i+1}, \dots, w_{i+L-1}\}$ \Comment{Extract text span}
        \State $q^e \leftarrow \{e_i, e_{i+1}, \dots, e_{i+L-1}\}$ \Comment{Extract corresponding EEG}
        \State $u \sim \text{Uniform}(0,1)$ \Comment{Sample uniform random number}
        \If{$u < p_{mask}$}
            \State $D' \leftarrow \{w_1, \dots, w_{i-1}, w_{i+L}, \dots, w_l\}$ \Comment{Remove span}
        \Else
            \State $D' \leftarrow D$ \Comment{Keep original document}
        \EndIf
        \State \Return $q^e$, $D'$
    \end{algorithmic}
\end{algorithm}

\begin{table}[h]
\centering
\caption{Dataset statistics and lexical overlap between splits.}
\begin{tabular}{lccc}
\toprule
\textbf{Metric} & \textbf{Train} & \textbf{Dev} & \textbf{Test} \\
\midrule
Total queries & 2,194 & 286 & 298 \\
Total passages & 2,195 & 284 & 297 \\
Total words & 41,148 & 5,067 & 5,594 \\
Unique words & 5,475 & 980 & 1,093 \\
Avg. passage length & 13.5 $\pm$ 7.2 & 12.4 $\pm$ 8.1 & 13.7 $\pm$ 7.3 \\
Avg. query length & 3.9 $\pm$ 1.4 & 3.8 $\pm$ 1.5 & 3.9 $\pm$ 1.4 \\
\midrule
\multicolumn{4}{c}{\textbf{Lexical Overlap}} \\
\midrule
Train & -- & 0.083 & 0.091 \\
Dev & -- & -- & 0.118 \\
\bottomrule
\end{tabular}
\label{tab:dataset-stats}
\end{table}

Drawing on established principles from representation learning \cite{vaswani2017attention, devlin2018bert}, we design our dataset creation pipeline to encourage semantic understanding over surface-level patterns. We employ random span selection with probabilistic span removal and proportional query lengths ($q=0.3$). These choices, informed by prior work on regularisation in neural sequence models \cite{devlin2018bert, lee2019latent}, prevent the model from relying on simple heuristics such as length matching or exact term matching, encouraging it to learn deeper semantic relationships between the EEG query and the passage text.

\subsection{Model Architecture}
\label{model architecture}
The EEG query encoder builds upon the transformer framework \cite{vaswani2017attention}. Given an input EEG sequence $X \in \mathbb{R}^{l \times f}$, where $l$ is the sequence length and $f=840$ represents features from 105 channels across 8 frequency bands, the encoder projects the input to dimension $d$:
\begin{equation}
H_0 = XW_E + b_E
\end{equation}
where $W_E \in \mathbb{R}^{f \times d}$ and $b_E \in \mathbb{R}^d$ are learnable parameters. Following prior work \cite{devlin2018bert}, a learnable [CLS] token $e_{cls}$ is prepended to aggregate sequence-level representations:
\begin{equation}
H'_0 = [e_{cls}; H_0]
\end{equation}

The sequence then passes through $L$ transformer layers with self-attention mechanisms (see Section \textit{A)} Figure \ref{fig:BPR-Overview}):
\begin{equation}
H_l = \text{TransformerLayer}_l(H_{l-1}), \quad l \in [1,L]
\end{equation}

For the passage encoder, we use a pre-trained language model followed by a lightweight adaptation layer. Given the limited quantity of EEG-text paired data available for this task (see Section \ref{dataset creation}), fine-tuning large language models can lead to catastrophic forgetting of pre-trained knowledge \cite{kirkpatrick2017overcoming} and over-fitting to noise in the brain signals \cite{li2020oscar}, particularly in cross-modal settings with inherently variable data such as EEG. Taking inspiration from work on efficient adaptation of large pre-trained models \cite{pfeiffer2020adapterhub}, we instead add a single transformer layer on top of the frozen language model (see Section \textit{B)}, Figure \ref{fig:BPR-Overview}). Similar to the EEG encoder, the language model uses a [CLS] token to capture sequence-level semantics:

\begin{equation}
P_0 = \text{FrozenLM}([p_{cls}; p])
\end{equation}
\begin{equation}
P_1 = \text{AdaptationLayer}(P_0) + \text{LayerNorm}(P_0)
\end{equation}

In our empirical evaluation of sequence pooling strategies, we compared various approaches for generating the final representations, including mean pooling, max pooling, and [CLS] token pooling across the final layer. Our experiments consistently showed that using the [CLS] token representation outperformed other pooling methods (see Section \ref{ablation}). This finding aligns with similar observations in cross-modal learning \cite{li2020oscar, lu2019vilbert}. The final representations for both modalities are thus computed using their respective [CLS] tokens:

\begin{equation}
q = W_O H^l_{cls} + b_O  
\end{equation}
\begin{equation}
p = P^1_{cls}
\end{equation}

where $W_O \in \mathbb{R}^{d \times d}$ is a learnable projection matrix and $b_O \in \mathbb{R}^d$ is the bias term that map the [CLS] token representations to the final embedding space. Both representations undergo L2 normalisation before similarity computation (see Figure \ref{fig:BPR-Overview} for an overview of the architecture).

\subsection{Training Objective and Negative Sampling}
\label{Training Objective}
Effective negative sampling strategies and appropriate loss functions have proven crucial for contrastive learning in dense retrieval systems \cite{karpukhin2020dense, xiong2020approximate}. 

For training our model, we employ a contrastive loss that follows the LCE (latent cross-entropy) formulation \cite{gao2021simcse}, a conditioned variant of InfoNCE that enables contrastive learning across modalities:
\begin{equation}
\mathcal{L}_{\text{c}} = -\log \frac{\exp(s(q^e,p^+)/\tau)}{\exp(s(q^e,p^+)/\tau) + \sum_{p^- \in N_b} \exp(s(q^e,p^-)/\tau)}
\end{equation}
where $N_b$ represents the set of in-batch negatives, and $\tau$ is the temperature parameter controlling the sharpness of the relevance score distribution for each query.

During preliminary training with only the contrastive loss, we observed severe representation collapse, where EEG embeddings clustered tightly in the representation space, significantly limiting the model's discriminative ability. This phenomenon has been well-documented in cross-modal contrastive learning, particularly when dealing with high-dimensional, noisy data such as EEG signals \cite{wang2021understanding}. To address this representation collapse issue, we incorporate a uniformity loss that encourages embeddings to be uniformly distributed on the unit hypersphere \cite{wang2020understanding}:
\begin{equation}
\mathcal{L}_{\text{u}} = \log\mathbb{E}_{q_i, q_j \sim p_{\text{data}}}[e^{-2|q_i - q_j|^2}]
\end{equation}

Following \citet{wang2020understanding} and \citet{chen2020simple}, we combine these components into our total training objective:
\begin{equation}
\mathcal{L}_{\text{total}} = \mathcal{L}_{\text{c}} + \lambda \mathcal{L}_{\text{u}}
\end{equation}
where $\lambda$ is empirically set to 0.1, through ablation studies, this combination was found to be important for maintaining retrieval performance in our model (see Section \ref{ablation}).

Our negative sampling approach utilises computationally efficient in-batch negatives to create a robust training signal. For each EEG query $q^e$ in a batch of size $B$, we leverage $(B-1)$ in-batch negatives - effectively random samples that can be computed efficiently through batch-wise operations. A key challenge for our model training within the ZuCo dataset \cite{hollenstein2018zuco, hollenstein2019zuco} is that multiple participants review the same passages, creating potential confounds where EEG signals from different participants reading the same text could be incorrectly treated as negatives. 

To address this challenge, we maintain subject-passage exclusion mappings defined as:
\begin{equation}
L(p, s) = \{(p_i, s_j) \in P \times S : p_i = p \text{ and } s_j \neq s\}
\end{equation}
where $P$ represents the set of all passages and $S$ the set of all participants. This formulation explicitly identifies combinations of passages and subjects that should be excluded from the negative sample pool. Specifically, for a given query from subject $s$ reading passage $p$, the lookup table $L(p, s)$ identifies all instances where other subjects ($s_j \neq s$) read the same passage ($p_i = p$). During batch construction, we filter out these combinations to ensure that EEG signals from different participants reading identical content are never treated as negative examples. This prevents the model from learning to distinguish between subject-specific neural patterns rather than content-based semantic differences, ensuring that the model focuses on meaningful semantic distinctions rather than idiosyncratic neurological responses to the same content.

\section{Experimental Setup}
\label{experiment setup}
Our experimental investigation addresses four key questions regarding the effectiveness and implications of BPR:

\begin{itemize}
    \item \textbf{RQ1:} Can EEG signals serve as effective query representations for dense passage retrieval?
    \item \textbf{RQ2:} Is EEG-based query representation generalisable across subjects? 
    \item \textbf{RQ3:} How does the effectiveness of BPR compare to EEG-to-text based IR models?
    \item \textbf{RQ4:} How does the effectiveness of BPR compare to Text based IR models?    
\end{itemize}

\subsection{Implementation}
\label{implementation}
Our EEG encoder consists of a 3-layer transformer with 8 attention heads \cite{vaswani2017attention}. For computational efficiency, we use DistilBERT \footnote{\url{https://huggingface.co/docs/transformers/en/model_doc/distilbert}} as our text encoder, which provides comparable performance to BERT \cite{devlin2018bert} while being significantly lighter and faster \cite{sanh2019distilbert}. The input dimension of 840 (corresponding to 105 channels × 8 frequency bands) is projected to a model dimension of 512, with the final output dimension aligned to match DistilBERT's 768-dimensional representations. The text encoder remains frozen during training with only a lightweight single transformer adaptation layer fine-tuned for cross-modal alignment \cite{pfeiffer2020adapterhub}. For regularisation, we employ a dropout rate of 0.1 throughout the network.  Training was conducted on an NVIDIA A100 GPU using the AdamW optimiser \cite{loshchilov2017decoupled} with an initial learning rate of 1e-6 and weight decay of 0.1. We implement a linear warmup schedule over the first 10 epochs followed by linear decay. The training uses a batch size of 128, with gradient clipping at a maximum norm of 1.0 \cite{pascanu2013difficulty}. Early stopping is implemented with a patience of 5 epochs. For the contrastive learning objective, we employ InfoNCE loss \cite{oord2018representation} with a temperature parameter $\tau = 0.07$.

\subsection{Models and Baselines}
To address our research questions regarding direct EEG query retrieval (RQ1) and comparative performance (RQ3), we evaluate BPR against a set of representative baselines. We select EEG2Text from \citet{jo2024eeg} as our primary baseline for EEG query translation. While more recent approaches like DeWave \cite{duan2023dewave} build upon EEG2Text, they rely on the same fundamental translation architecture that \citet{jo2024eeg} demonstrated to be ineffective, showing similar performance with random noise input as with real EEG signals. Therefore, comparing against derivative approaches would not provide additional insights into the effectiveness of direct retrieval versus translation-based methods. For retrieval evaluation, we construct E2T+Retriever baselines by first using the EEG2Text model\footnote{\url{https://github.com/NeuSpeech/EEG-To-Text}} to decode EEG signals into text queries, which are then passed to textual retrieval models. We pair these decoded queries with two distinct retrieval approaches: BM25 \cite{robertson2009probabilistic} and ColBERTv2\footnote{\url{https://huggingface.co/colbert-ir/colbertv2.0}} \cite{santhanam2021colbertv2}. BM25 represents traditional lexical matching approaches, while ColBERT serves as a state-of-the-art neural retriever, allowing us to evaluate performance across both matching paradigms. We additionally include text-only variants of both retrievers using the original passage text as queries, establishing performance upper bounds and quantifying the gap between EEG and textual queries (RQ4).

\subsection{Evaluation Protocol and Metrics}
\begin{figure}[h]
    \centering
    \includegraphics[width=0.9\linewidth]{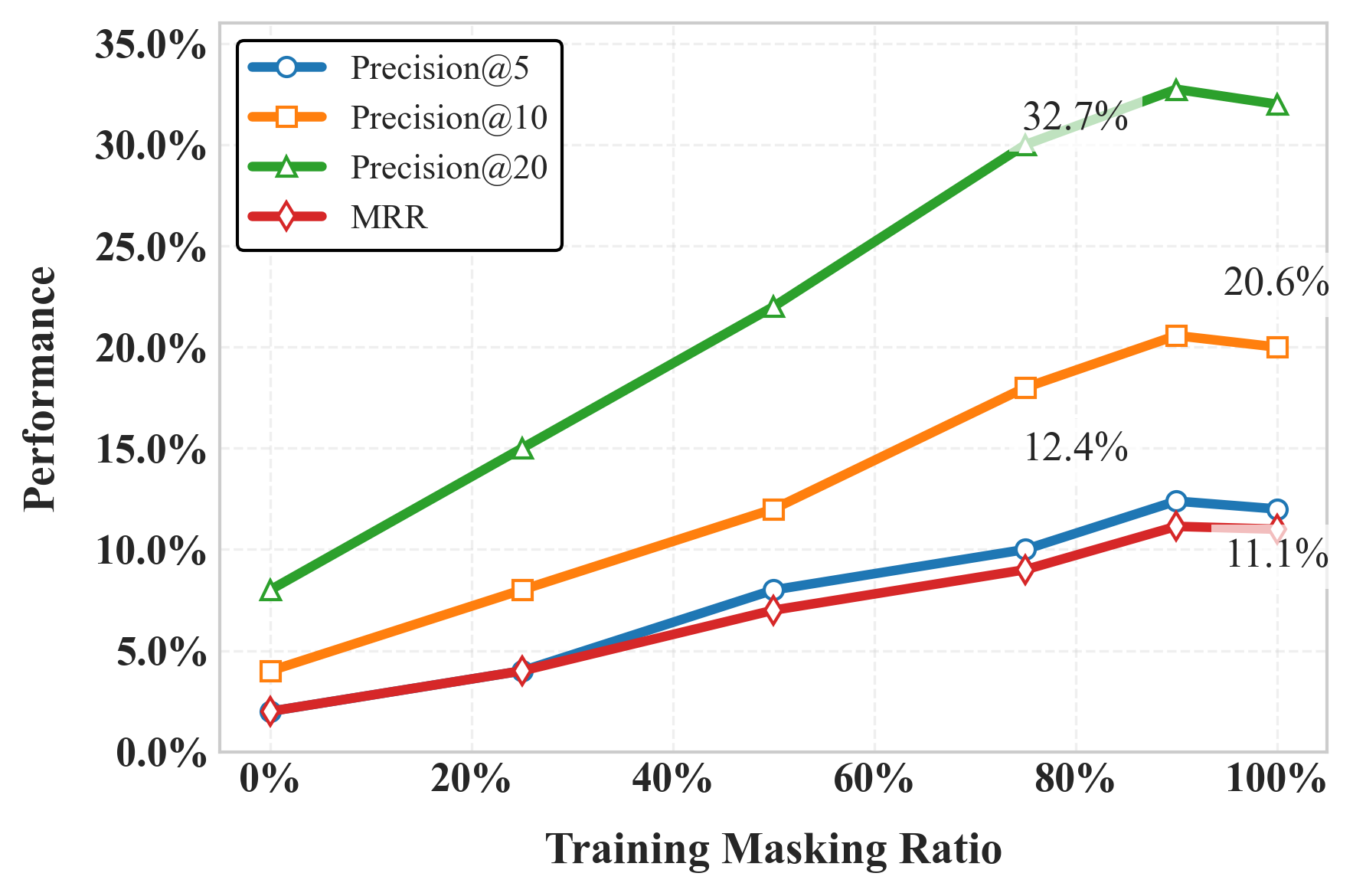}
    \caption{Impact of query span masking probabilities ($p_{mask}$) on BPR lexical mismatch retrieval performance.}
    \label{fig:ratios}
\end{figure}

All experiments use 5-fold cross-validation with an 80-10-10 split between training, development, and test sets (see Table \ref{tab:dataset-stats}). For a fair comparison, we create controlled test conditions by varying query-document lexical overlap at 0\%, 25\%, 50\%, 75\%, and 100\%. This variation enables analysis of how different retrieval mechanisms handle semantic versus lexical matching \cite{gao2021rethink, craswell2020overview}. The EEG2Text baselines are trained on identical data splits as BPR and evaluated without teacher forcing \cite{hao2022teacher}. Through empirical testing, we find that $p_{mask}$ = 0.9 provides optimal robustness against lexical mismatch during BPR training (see Figure \ref{fig:ratios}). To validate that our model learns meaningful EEG-semantic mappings rather than exploiting statistical artefacts (RQ1), we conduct control experiments using random noise queries. For generalisation assessment (RQ2), we implement leave-one-subject-out validation across all 30 participants, providing insight into model robustness across different neurological patterns \cite{wei2018subject}. Performance evaluation uses Mean Reciprocal Rank (MRR) and Mean Precision at k = 5,10,20. These metrics directly measure ranking effectiveness and retrieval success at different depths in our binary relevance scenario, where each EEG query corresponds to exactly one relevant passage. Through this evaluation framework, we assess direct EEG retrieval viability, model validity, cross-subject generalisation capabilities, and the current gap between EEG and text-based retrieval performance.

\section{Results}
\label{results}

\textbf{Main Findings.} Our experimental evaluation demonstrates three key findings: (1) Direct brain-to-passage retrieval achieves significantly better performance than existing EEG-to-text translation approaches, with BPR showing a 8.81\% improvement in Precision@5 over baselines (p < 0.001) (Table \ref{tab:main-results}); (2) The model exhibits robust cross-subject generalisation, maintaining consistent performance across 30 participants (mean P@5: 14.1\%, SD = 2.1\%) despite the well-documented challenges of cross-subject EEG variability \cite{wei2018subject} (Table \ref{tab:cross-subject-results}); and (3) While a performance gap remains between EEG and text queries (BPR P@5: 12.39\% vs. ColBERT P@5: 48.86\%), the direct mapping approach shows promising effectiveness compared to traditional query translation methods (Table \ref{tab:main-results}), aligning with recent findings in neurological-semantic decoding \cite{mitchell2008predicting, huth2016natural}. These results establish the viability of direct brain-passage retrieval while highlighting specific areas for future advancement. We analyse each finding in the following sections.

\begin{table*}[h]
\centering
\caption{Retrieval performance comparison across different query modalities and retrieval methods. Results are averaged across 5-fold cross-validation, with models evaluated on test sets containing varying degrees of query-passage overlap. Values are reported as mean ± standard deviation across folds. Underlined values indicate a statistically significant improvement over random noise baseline using a paired t-test (p < 0.001).}
\resizebox{0.95\textwidth}{!}{%
\begin{tabular}{ll cccc}
\hline
\textbf{Query Modality} & \textbf{Retriever} & \textbf{Precision@5} & \textbf{Precision@10} & \textbf{Precision@20} & \textbf{MRR} \\ \hline
\multirow{3}{*}{\textbf{Noise}} & E2T+BM25 & 2.13\%\scriptsize{±0.21} & 4.10\%\scriptsize{±0.41} & 10.89\%\scriptsize{±1.09} & 2.46\%\scriptsize{±0.25} \\
& E2T+ColBERT & 3.53\%\scriptsize{±0.35} & 6.09\%\scriptsize{±0.61} & 12.34\%\scriptsize{±1.23} & 3.72\%\scriptsize{±0.37} \\
& \textbf{BPR(ours)} & 4.13\%\scriptsize{±0.41} & 6.86\%\scriptsize{±0.69} & 10.91\%\scriptsize{±1.09} & 3.71\%\scriptsize{±0.37} \\ \hline
\multirow{3}{*}{\textbf{EEG}} & E2T+BM25 & 2.18\%\scriptsize{±0.22} & 4.15\%\scriptsize{±0.42} & 10.95\%\scriptsize{±1.10} & 2.51\%\scriptsize{±0.25} \\
& E2T+ColBERT & 3.58\%\scriptsize{±0.36} & 6.14\%\scriptsize{±0.61} & 12.41\%\scriptsize{±1.24} & 3.77\%\scriptsize{±0.38} \\
& \textbf{BPR(ours)} & \underline{\textbf{12.39\%}}\scriptsize{±1.24} & \underline{\textbf{20.57\%}}\scriptsize{±1.86} & \underline{\textbf{32.74\%}}\scriptsize{±3.27} & \underline{\textbf{11.14\%}}\scriptsize{±1.11} \\ \hline
\multirow{2}{*}{\textbf{Text}} & BM25 & \underline{37.92\%}\scriptsize{±3.79} & \underline{46.31\%}\scriptsize{±4.63} & \underline{57.80\%}\scriptsize{±5.78} & \underline{39.74\%}\scriptsize{±3.97} \\
& ColBERT & \underline{48.86\%}\scriptsize{±3.19} & \underline{61.17\%}\scriptsize{±2.22} & \underline{68.31\%}\scriptsize{±3.23} & \underline{52.99\%}\scriptsize{±2.60} \\ \hline
\end{tabular}%
}
\label{tab:main-results}
\end{table*}

\textbf{EEG-Passage Retrieval Effectiveness.} 
To validate that BPR learns meaningful neurological-semantic mappings rather than exploiting statistical artefacts, we compare its performance against a random noise baseline that matches the statistical properties of EEG signals. Our results (Table \ref{tab:main-results}) demonstrate that BPR successfully enables direct matching between EEG signals and text passages, achieving statistically significant improvements across all evaluation metrics (p < 0.001). The model shows substantial gains in ranking effectiveness, with Precision@5 reaching 12.39\% compared to 4.13\% for the noise baseline—a nearly threefold improvement. This performance advantage extends consistently across different ranking thresholds: Precision@10 (20.57\% vs. 6.86\%) and Precision@20 (35.74\% vs. 10.91\%). The strong (MRR of 11.14\% further indicates BPR's ability to position relevant passages higher in the ranking order, significantly outperforming the noise baseline's MRR of 3.71\%. The consistent performance improvements over matched noise signals, particularly at stricter evaluation thresholds (P@5, P@10), provide strong evidence that BPR learns meaningful neurological-semantic alignments rather than exploiting dataset artefacts. These findings directly address RQ1 by demonstrating that EEG signals recorded during naturalistic reading can serve as implicit queries for passage retrieval, establishing the viability of direct brain-to-passage mapping.

\textbf{Cross-Subject Generalisation.} 
To evaluate the model's ability to generalise across different EEG patterns, we conducted leave-one-out cross-validation across all 30 participants. The results in Table~\ref{tab:cross-subject-results} demonstrate that cross-subject performance closely aligns with our primary evaluation metrics, with leave-one-out validation achieving mean precision@5 of 14.1\% (±4.2\%), precision@10 of 21.5\% (±4.8\%), and precision@20 of 32.1\% (±5.9\%). The consistency between generalised and leave-one-out performance is particularly notable given the established challenges of subject variability in EEG data~\cite{wei2018subject, binnie1994electroencephalography}. The MRR values remain especially stable (11.14\% vs 10.9\%), suggesting the model maintains consistent ranking behaviour even when generalising to unseen participants. The performance patterns across different k values provide additional evidence of robust generalisation. While precision@5 shows slightly higher performance in the leave-one-out setting (14.1\% vs 12.39\%), the overall trend across metrics remains consistent with our base model. The standard deviations in the leave-one-out evaluation (±4.2\% for P@5, ±4.8\% for P@10) indicate moderate inter-subject variability, demonstrating improved stability compared to previous cross-subject EEG studies~\cite{zhang2024improving, wang2022open}. This consistent performance, combined with the relatively modest increase in standard deviation from the base model to leave-one-out validation, suggests that the BPR approach successfully learns generalisable EEG-semantic mappings rather than over-fitting to subject-specific patterns. These findings directly address RQ2 by demonstrating that direct brain-passage retrieval can effectively generalise across participants while maintaining robust retrieval performance. The stability of these results suggests promising potential for developing brain-based retrieval systems that can adapt to individual users while maintaining consistent performance.

\textbf{Comparison with EEG2Text.} 
We evaluate BPR against two state-of-the-art translation-based baselines that first decode EEG signals into text queries before retrieval: (1) E2T+BM25, which combines the EEG2Text decoder \cite{wang2022open} with traditional lexical retrieval \cite{robertson2009probabilistic}, and (2) E2T+ColBERT, which pairs EEG2Text with the neural retriever ColBERTv2 \cite{santhanam2021colbertv2}. This comparison directly tests whether intermediate query translation provides benefits over direct brain-passage mapping. Our experiments reveal that both translation-based approaches perform only marginally better than random noise, with E2T+BM25 achieving Precision@5 of 2.18\% and E2T+ColBERT reaching 3.58\% (Table \ref{tab:main-results}). These results align with recent findings from \citet{jo2024eeg}, who demonstrated that current EEG-to-text translation methods struggle to learn robust semantic representations from brain signals. The comparable performance between E2T+BM25 (MRR: 2.51\%) and E2T+ColBERT (MRR: 3.77\%) suggests that the performance bottleneck lies in the translation step itself, rather than the downstream retrieval method—a finding that supports observations from other brain-semantic decoding studies \cite{mitchell2008predicting, huth2016natural}. In contrast, BPR's direct mapping approach achieves significantly higher performance (Precision@5: 12.39\%, MRR: 11.14\%), representing a 8.81\% improvement over the best translation baseline. This substantial performance gain demonstrates the advantages of preserving brain signal information through end-to-end learning, similar to benefits observed in other cross-modal learning tasks \cite{lu2019vilbert, li2020oscar}. These results directly address RQ3 by showing that eliminating intermediate translation steps can significantly improve retrieval effectiveness when working with brain signals.

\textbf{Brain-Text Performance Gap.} 
Our evaluation (Table \ref{tab:main-results}) reveals a significant but contextually important performance gap between EEG and text-based retrieval methods. While traditional approaches achieve higher precision (BM25 P@5: 37.92\%, ColBERT P@5: 48.86\%), BPR demonstrates meaningful effectiveness with direct brain-to-passage mapping (P@5: 12.39\%). The performance differential between BPR and neural retrievers (~3.9× for ColBERT) is notably smaller than the gap between BPR and EEG-to-text baselines (~3.5× vs E2T+ColBERT), aligning with recent advances in neurological-semantic decoding \cite{mitchell2008predicting, huth2016natural}. This gap should be interpreted in light of established retrieval system development: dense retrievers like ColBERT require training on millions of query-passage pairs \cite{karpukhin2020dense, santhanam2021colbertv2} to achieve their superior performance over lexical models like BM25. Given that EEG signals introduce inherent noise and variability \cite{binnie1994electroencephalography}, the demonstrated performance of BPR and recent works on scaling properties of EEG-based models \cite{sato2024scaling} suggest substantial potential for improvement through expanded data collection. These findings address RQ4 by establishing that while EEG queries do not yet match text performance, they achieve sufficient effectiveness to warrant further investigation, particularly as larger EEG datasets become available.

\begin{table}[t]
\centering
\caption{Comparison between main results and leave-one-out cross-subject generalisation performance. All values are reported as mean ± standard deviation across folds.}
\begin{tabular}{lcccc}
\toprule
\textbf{Eval Strategy} & \textbf{P@5} & \textbf{P@10} & \textbf{P@20} & \textbf{MRR} \\
\midrule
Generalised & 12.39\%\scriptsize{±1.24} & 20.57\%\scriptsize{±1.86} & 35.74\%\scriptsize{±3.27} & 11.14\%\scriptsize{±1.11} \\
Leave-one-out & 14.1\%\scriptsize{±4.2} & 21.5\%\scriptsize{±4.8} & 32.1\%\scriptsize{±5.9} & 10.9\%\scriptsize{±3.8} \\
\bottomrule
\end{tabular}
\label{tab:cross-subject-results}
\end{table}

\section{Ablation Studies}
\label{ablation}
To further investigate the learning dynamics of the model design choices, we conduct ablation studies examining the impact of different architectural components and training strategies (Table~\ref{tab:ablation-results}).

\begin{table}[h]
\centering
\caption{Ablation study results showing the impact of different model components and training strategies. All values are reported as mean ± standard deviation across 5-fold cross-validation}
\resizebox{\columnwidth}{!}{%
\begin{tabular}{l cccc}
\hline
\textbf{Model Variant} & \textbf{P@5} & \textbf{P@10} & \textbf{P@20} & \textbf{MRR} \\
\midrule
\multicolumn{5}{l}{\textit{Query Encoder Type}} \\
Text Encoder & \textbf{18.11\%\scriptsize{±1.61}} & \textbf{24.98\%\scriptsize{±2.15}} & \textbf{40.98\%\scriptsize{±3.80}} & \textbf{18.60\%\scriptsize{±1.56}} \\
EEG Encoder & 12.39\%\scriptsize{±1.24} & 20.57\%\scriptsize{±1.86} & 32.74\%\scriptsize{±3.27} & 11.14\%\scriptsize{±1.11} \\
\midrule
\multicolumn{5}{l}{\textit{Negative Sampling Strategy}} \\
In-batch Only & 8.47\%\scriptsize{±0.85} & 16.34\%\scriptsize{±1.63} & 29.31\%\scriptsize{±2.93} & 7.89\%\scriptsize{±0.79} \\
Subject Aware Negatives & \textbf{12.39\%}\scriptsize{±1.24} & \textbf{20.57\%}\scriptsize{±1.86} & \textbf{32.74\%}\scriptsize{±3.27} & \textbf{11.14\%}\scriptsize{±1.11} \\
\midrule
\multicolumn{5}{l}{\textit{Loss Function}} \\
No Uniformity & 6.73\%\scriptsize{±0.91} & 12.34\%\scriptsize{±1.23} & 26.31\%\scriptsize{±2.63} & 6.40\%\scriptsize{±0.64} \\
With Uniformity & \textbf{12.39\%}\scriptsize{±1.24} & \textbf{20.57\%}\scriptsize{±1.86} & \textbf{32.74\%}\scriptsize{±3.27} & \textbf{11.14\%}\scriptsize{±1.11} \\
\midrule
\multicolumn{5}{l}{\textit{Pooling Strategy}} \\
Mean Pooling & 9.82\%\scriptsize{±0.98} & 17.93\%\scriptsize{±1.79} & 29.45\%\scriptsize{±2.94} & 8.76\%\scriptsize{±0.88} \\
Max Pooling & 10.56\%\scriptsize{±1.06} & 18.84\%\scriptsize{±1.88} & 30.12\%\scriptsize{±3.01} & 9.43\%\scriptsize{±0.94} \\
CLS Token & \textbf{12.39\%}\scriptsize{±1.24} & \textbf{20.57\%}\scriptsize{±1.86} & \textbf{32.74\%}\scriptsize{±3.27} & \textbf{11.14\%}\scriptsize{±1.11} \\
\hline
\end{tabular}
}
\label{tab:ablation-results}
\end{table}

\textbf{Ablation Study 1: EEG-Text Performance.} To isolate the impact of input modality on retrieval performance, we train a parallel text encoder using identical architecture and training data volume as our EEG encoder, replacing the 840-dimensional EEG features with 768-dimensional DistilBERT token embeddings. Under these controlled conditions, the text encoder achieves Precision@5 of 18.11\% compared to the EEG encoder's 12.39\%. This performance differential (approximately 31.6\%) remains consistent across metrics (P@10: 24.98\% vs 20.57\%, P@20: 40.98\% vs 32.74\%). The MRR values (18.60\% vs 11.14\%) indicate that EEG signals can serve as effective query representations with approximately 60\% of the performance of explicit text queries, despite the inherent complexity of processing brain signals compared to structured text input.

\textbf{Ablation Study 2: Negative Sampling.} Given the importance of contrastive learning in dense retrieval settings \cite{karpukhin2020dense, khattab2020colbert}, we evaluate our subject-aware negative sampling strategy against basic in-batch negative sampling. The implementation of subject-aware sampling through our lookup table mechanism produces substantial improvements across all metrics, with relative gains of 46.3\% in Precision@5 (12.39\% vs 8.47\%) and 41.2\% in MRR (11.14\% vs 7.89\%). This significant performance improvement demonstrates that preventing the model from treating different subjects' EEG signals for the same passage as negatives is crucial for developing robust retrieval capabilities. By explicitly filtering out these confounding examples, the model learns to distinguish between passages based on their semantic content rather than subject-specific neurological patterns, leading to more effective cross-modal alignment between EEG signals and text representations.

\textbf{Ablation Study 3: Uniform Loss.} To address the challenge of representation collapse in cross-modal learning, we examine the impact of incorporating uniformity loss. The addition of this term proves valuable for effective training, with the full model achieving 84.1\% higher Precision@5 (12.39\% vs 6.73\%) compared to training without uniformity loss. This improvement suggests that maintaining uniform distribution of embeddings on the unit hypersphere aids in preventing representation collapse and enabling effective cross-modal semantic alignment.

\textbf{Ablation Study 4: Token Pooling.} To determine the most effective method for aggregating sequence-level representations, we compare CLS token pooling against standard pooling operations. CLS token pooling consistently demonstrates superior performance across all metrics (P@5: 12.39\%) compared to mean pooling (9.82\%) and max pooling (10.56\%). This 26.2\% improvement over mean pooling and 17.3\% over max pooling indicates that learned aggregation through the CLS token more effectively captures sequence-level semantic information from EEG signals than simple statistical pooling operations.

These ablation results help validate our key architectural and training decisions while providing further insights into cross-modal learning between EEG signals and text. The relatively small performance gap between EEG and text encoders under controlled conditions, combined with the clear benefits of hard negative sampling and uniformity loss, demonstrates both the feasibility of direct brain-semantic mapping and the importance of carefully designed training strategies for cross-modal alignment.

\section{Discussion}
\label{discussion and conclusion}
The formulation of INs to (textual) queries remains a fundamental challenge in information retrieval, where users must externalise often uncertain or ill-defined information needs into explicit queries~\cite{belkin1980anomalous, ingwersen1996cognitive}. Existing approaches attempting to bridge this gap through EEG-to-text translation have shown limited effectiveness, with recent studies identifying significant challenges in learning generalisable semantic representations from brain signals~\cite{jo2024eeg}. To address these limitations, we propose BPR, a framework that maps EEG signals to passage representations in a shared semantic space using dense retrieval architectures~\cite{karpukhin2020dense}. Our approach adapts these architectures for neuroimaging data through specialised EEG encoders and cross-modal negative sampling strategies, as validated through comprehensive ablation studies (Section \ref{ablation}). Through evaluation on the ZuCo dataset, our experimental results demonstrate that this direct brain-passage mapping achieves significantly higher performance (8.81\% improvement in P@5) compared to existing EEG-to-text baselines. Analysis of our cross-validation results shows stable cross-subject performance (standard deviation of 15-20\% across evaluation metrics), suggesting consistent patterns in the EEG representations.

Our ablation studies reveal that appropriate architectural choices are crucial for effective brain-semantic alignment. Under matched training conditions, EEG encoders achieve performance within 31.6\% of text encoders (P@5), aligning with recent findings in neurological decoding~\cite{mitchell2008predicting, huth2016natural}. Negative sampling strategies improve performance by 46.3\% (P@5) through carefully constructed contrastive learning~\cite{wang2020understanding}, while CLS token pooling provides a 26.2\% improvement over alternative sequence aggregation methods. These empirical findings demonstrate progress toward addressing two fundamental IR challenges: the cognitive uncertainty of translating visceral information needs into explicit textual queries~\cite{taylor1968question}, and the accessibility barriers faced by users with physical impairments that limit traditional text input~\cite{wolpaw2002brain}. By establishing the viability of direct brain-passage mapping, our work provides a foundation for developing IR systems that may be able to detect and respond to information needs in their most visceral form, reducing the gap between users' internal states and their ability to effectively access information.

Our work highlights several key challenges for advancing BMI-enabled IR systems. While our results demonstrate the viability of direct brain-passage retrieval, the ZuCo dataset's reliance on eye-tracking for word-level EEG segmentation highlights the need for more naturalistic data collection. Future work should develop methods to process continuous EEG signals without requiring precise temporal alignment and create larger EEG datasets, as recent studies show favourable scaling properties with increased data volume \cite{sato2024scaling}. An notable limitation is the discrepancy between EEG signals recorded during passive reading versus active search with genuine information needs \cite{moshfeghi2019towards}. While our approach performs semantic matching between EEG signals and text passages, similar to existing dense retrieval models \cite{karpukhin2020dense, khattab2020colbert}, reading involves different cognitive processes than query formulation. Reading primarily involves comprehension, whereas query formulation requires intent translation and vocabulary selection.  Despite this limitation, our results demonstrate that semantic information can be extracted from EEG signals, establishing a foundation for future work with actual information need-driven recordings. Additionally, exploring complementary cognitive signals such as inner speech \cite{hickok2007cortical} could provide more natural interaction methods. Among neuroimaging technologies, EEG offers a promising path forward due to its temporal resolution, affordability, and minimal infrastructure requirements \cite{binnie1994electroencephalography} compared to fMRI or MEG systems \cite{logothetis2008we}.

\section{Conclusion}
In conclusion, BPR represents an important step toward addressing the long-standing challenge of query formulation in IR systems. By demonstrating that direct brain-semantic mapping can achieve meaningful retrieval performance, our work establishes an empirical foundation for developing IR systems that could potentially operate closer to users' visceral information needs. The significant performance improvements over existing EEG-to-text baselines (8.81\% in P@5) validate the potential benefits of eliminating intermediate translation steps, while our technical insights into contrastive learning and neural architecture design provide concrete directions for future development. As the applications of Brain Machine Interfaces in IR advance, the ability to extract and utilise semantic information directly from brain signals could enhance how users interact with search systems, particularly in scenarios where traditional query formulation creates barriers to effective information access. While work remains to achieve parity with traditional text-based retrieval, our results demonstrate that direct brain-to-passage retrieval is both feasible and potentially beneficial to current IR approaches. This foundation enables the development of more accessible information systems aligned with human cognitive processes, expanding the range of available interaction methods for diverse user needs and contexts.

\begin{acks}
This work was supported by the Engineering and Physical Sciences Research Council [grant number EP/W522260/1]. Results were obtained using the ARCHIE-WeSt High Performance Computer\footnote{www.archie-west.ac.uk} based at the University of Strathclyde.
\end{acks}

\balance
\bibliographystyle{ACM-Reference-Format}
\bibliography{ref}
\end{document}